\documentclass[twocolumn,preprintnumbers]{revtex4}%
\usepackage{amssymb}
\usepackage{amsmath}
\usepackage{graphicx}
\usepackage{epstopdf}
\usepackage{dcolumn}
\usepackage{bm}
\usepackage{xcolor}
\usepackage{amsfonts}
\usepackage[english]{babel}%
\begin{document}
\title{Polarization-selective magneto-optical modulation}
\author{Banoj Kumar Nayak}
\affiliation{Andrew and Erna Viterbi Department of Electrical Engineering, Technion, Haifa
32000 Israel}
\author{Eyal Buks}
\affiliation{Andrew and Erna Viterbi Department of Electrical Engineering, Technion, Haifa
32000 Israel}
\date{\today }

\begin{abstract}
We study magneto-optical coupling in a ferrimagnetic sphere resonator made of
Yttrium iron garnet. We find that the resonator can be operated in the telecom
band as a polarization-selective optical modulator. Intermodulation gain can
be employed in the nonlinear regime for amplification.

\end{abstract}
\pacs{}
\maketitle

\section{Introduction}

Information is commonly transmitted by modulating a monochromatic carrier
wave. The method of single sideband modulation (SSM) allows reducing both
transmission power and bandwidth, in comparison with simpler methods such as
amplitude, frequency and phase modulation \cite{lazzarini2008asymmetric}. In
the radio frequency band SSM can be implemented using electronic circuits,
however, SSM implementation in the optical band is challenging, since it
requires that different out of phase modulation methods are simultaneously
applied \cite{Li_1,Shimotsu_364}.

Magneto-optical (MO) coupling
\cite{Rameshti_1,Kusminskiy_299,Zhu_2012_11119,Juraschek_094407,Bittencourt_014409,Zhang_123605,
Stancil_Spin} in ferrimagnetic sphere resonators (FSR) can be used for optical
modulation of signals in the microwave band. Such a modulation has been
demonstrated before
\cite{Haigh_133602,Osada_103018,Osada_223601,Sharma_094412,Almpanis_184406,Zivieri_165406,Desormiere_379,Liu_3698,Chai_820,Zhu_1291,Li_040344}
by exciting individual whispering gallery FSR optical modes using either a
tapered optical fiber or a prism. Here we employed a modified experimental
setup, in which light in the telecom band is transmitted through the FSR bulk.
Driving the FSR near its resonance generates sidebands in the transmitted
optical spectrum. We find that the FSR can be used as a polarization-selective
SSM. The polarization selectivity is attributed to angular momentum
conservation in photon-magnon scattering
\cite{Wettling_211,cottam1986light,Liu_060405,Haigh_143601,Hisatomi_207401,Hisatomi_174427}%
. We demonstrate that intermodulation (IMD) gain can be exploited in the
nonlinear regime for amplification.

\section{Experimental setup}

\begin{figure}[ptb]
\begin{center}
\includegraphics[width=3.2in,keepaspectratio]{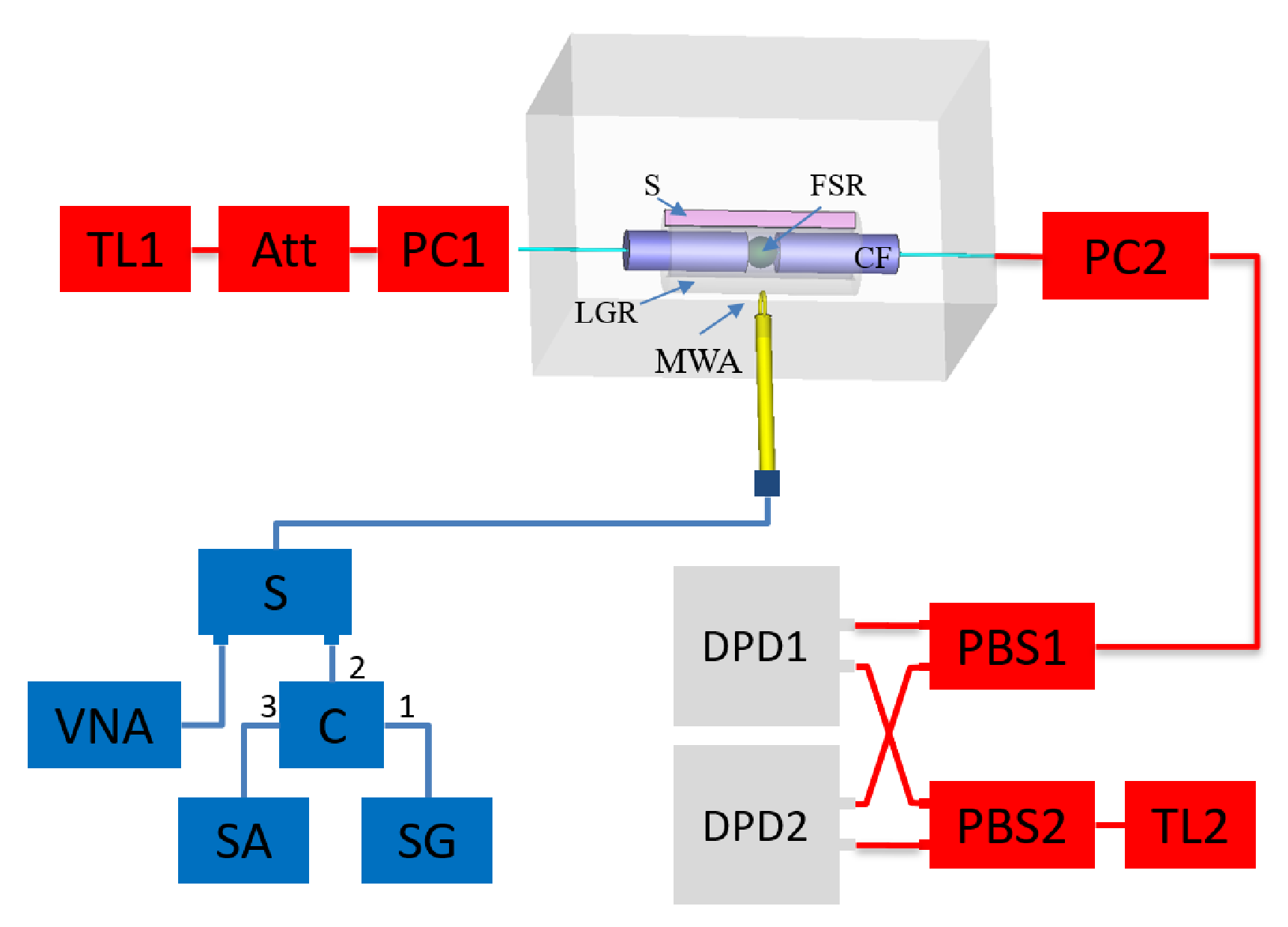}
\end{center}
\caption{{}Experimental setup. Optical fibers are installed on both sides of
the FSR for transmission of light through the sphere. Optical components [TL
(tunable laser), Att (optical attenuator), PC (polarization controller) and
PBS (polarization beam splitter)] and fibers are red colored, and MW
components [MWA (microwave loop antenna), S (splitter), C (circulator), VNA
(vector network analyzer), SA (spectrum analyzer) and SG (signal generator)]
and coaxial cables are blue colored. TL2 together with two PBSs (labelled as
PBS1 and PBS2) and two differential photo detectors (labelled as DPD1 and
DPD2) operate as a polarization-selective optical spectrum analyzer (OSA)
\cite{Baney_355}. A power amplifier is serially connected to the SG. The MWA
is weakly coupled to the FSR-LGR system.}%
\label{FigExSetup}%
\end{figure}

The experimental setup is schematically shown in Fig.~\ref{FigExSetup}.
Optical components and fibers are red colored, whereas blue color is used to
label microwave (MW) components and coaxial cables. A MW cavity made of a loop
gap resonator (LGR) allows achieving a relatively large coupling between
magnons and MW photons
\cite{Goryachev_054002,Zhang_205003,Mathai_054428,Nayak_062404}. The LGR is
fabricated from a hollow concentric aluminium tube. A sapphire (S) strip of
$260%
\operatorname{\mu m}%
$ thickness is inserted into the gap in order to increase its capacitance,
which in turn reduces the frequency $f_{\mathrm{c}}$ of the LGR fundamental
mode. An FSR made of Yttrium iron garnet (YIG) having radius of $R_{\mathrm{s}%
}=125%
\operatorname{\mu m}%
$ is held by two ceramic ferrules (CF) inside the LGR. The two CFs, which are
held by a concentric sleeve, provide transverse alignment for both input and
output single mode optical fibers. Fiber longitudinal alignment is performed by
maximizing optical transmission.

The angular frequency of the Kittel mode $\omega_{\mathrm{m}}$ is
approximately given by $\omega_{\mathrm{m}}=\mu_{0}\gamma_{\mathrm{e}%
}H_{\mathrm{s}}$, where $H_{\mathrm{s}}$ is the static magnetic field,
$\mu_{0}$ is the free space permeability, and $\gamma_{\mathrm{e}}/2\pi=28%
\operatorname{GHz}%
\operatorname{T}%
^{-1}$\ is the gyromagnetic ratio \cite{Stancil_Spin}. The applied static
magnetic field $\mathbf{H}_{\mathrm{s}}$ is controlled by adjusting the
relative position of a magnetized Neodymium using a motorized stage. The
static magnetic field is normal to the light propagation direction
$\mathbf{k}$, and the magnetic field of MW drive is nearly parallel to
$\mathbf{k}$. The LGR-FSR coupled system is encapsulated inside a metallic
rectangular shield made of aluminum. The LGR is weakly coupled to a microwave
loop antenna (MWA).

\begin{figure}[ptb]
\begin{center}
\includegraphics[width=2.6in,keepaspectratio]{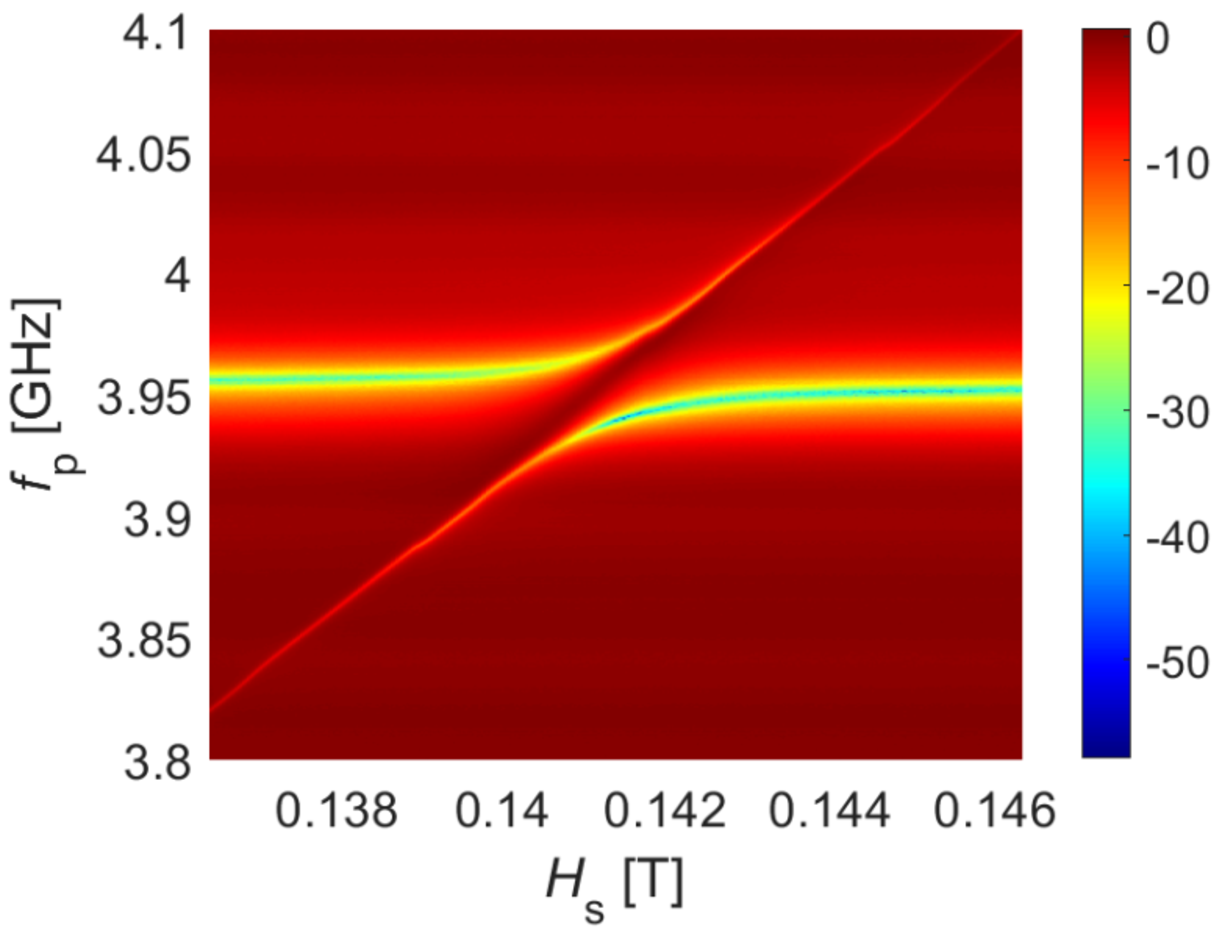}
\end{center}
\caption{{}VNA reflectivity in dB units as a function of magnetic field
$H_{\mathrm{s}}$ at applied microwave power of $-30$ dBm.}%
\label{FigVNAPcolor}%
\end{figure}

The plot in Fig.~\ref{FigVNAPcolor} exhibits a vector network analyzer\ (VNA)
reflectivity measurement of the LGR-FSR coupled system. The static applied
magnetic field $H_{\mathrm{s}}$ in this measurement is varied near the value
corresponding to avoided-crossing between the FSR and LGR resonances.

\section{Optical side bands}

\begin{figure}[ptb]
\begin{center}
\includegraphics[width=2.6in,keepaspectratio]{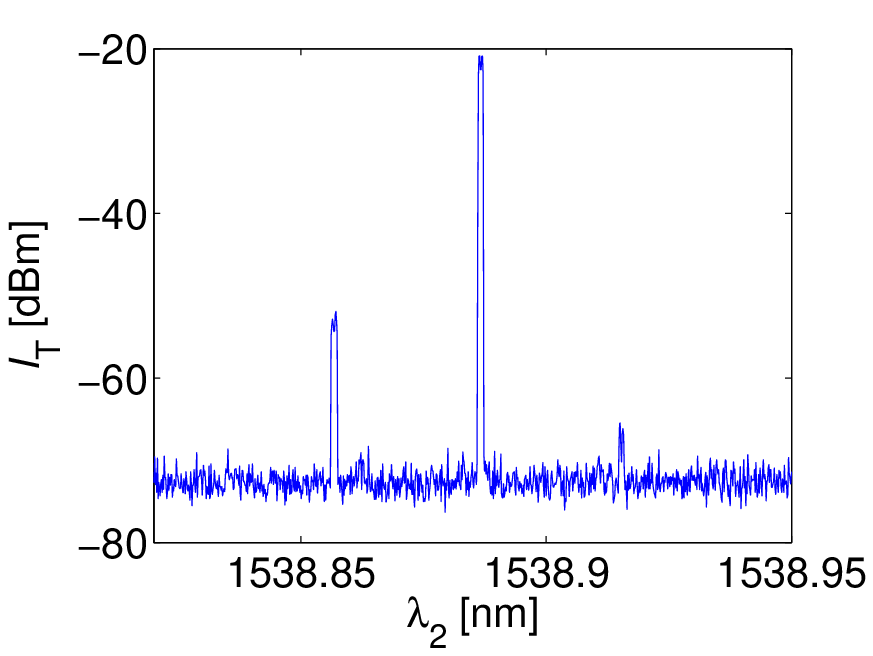}
\end{center}
\caption{The transmitted optical spectrum. For this measurement the TL1 is set
at optical power of $31\operatorname{mW}$ and wavelength $\lambda_{\mathrm{L}%
}$ of $1538.887\operatorname{nm}$, and the driving microwave is set at
frequency $\omega_{\mathrm{p}}/\left(  2\pi\right)  $ of
$3.79\operatorname{GHz}$ and power of $P_{\mathrm{p}}$ of $20$ dBm.}%
\label{FigOSAT}%
\end{figure}

Optical side bands are observed in the transmission spectrum when the driving
microwave frequency $\omega_{\mathrm{p}}/\left(  2\pi\right)  $ is tuned close
to the FSR resonance at $\omega_{\mathrm{m}}/\left(  2\pi\right)  $. The plot
shown in Fig.~\ref{FigOSAT} exhibits the measured total optical intensity
$I_{\mathrm{T}}=I_{\mathrm{DPD1}}+I_{\mathrm{DPD2}}$ as a function of the
wavelength $\lambda_{2}$ of TL2, where $I_{\mathrm{DPD1}}$ and
$I_{\mathrm{DPD2}}$ are the intensities measured by the two differential
photodetectors (labelled as DPD1 and DPD2 in Fig.~\ref{FigExSetup}). The side
band wavelengths are given by $\lambda_{\mathrm{L}}\pm\lambda_{\mathrm{SB}}$,
where $\lambda_{\mathrm{SB}}\simeq\lambda_{\mathrm{L}}^{2}\omega_{\mathrm{p}%
}/\left(  2\pi c\right)  $, and $\lambda_{\mathrm{L}}$ is the TL1 wavelength,
which is related to the TL1 frequency $\omega_{\mathrm{L}}/\left(
2\pi\right)  $ by $\omega_{\mathrm{L}}=2\pi c/\lambda_{\mathrm{L}}$, where $c$
is the speed of light in vacuum. The value of $\lambda_{\mathrm{SB}}=30.0%
\operatorname{pm}%
$ is obtained for TL1 wavelength of $\lambda_{\mathrm{L}}=1539%
\operatorname{nm}%
$ and FSR driving frequency of $\omega_{\mathrm{p}}/\left(  2\pi\right)  =3.79%
\operatorname{GHz}%
$.

\begin{figure}[b]
\begin{center}
\includegraphics[width=3.2in,keepaspectratio]{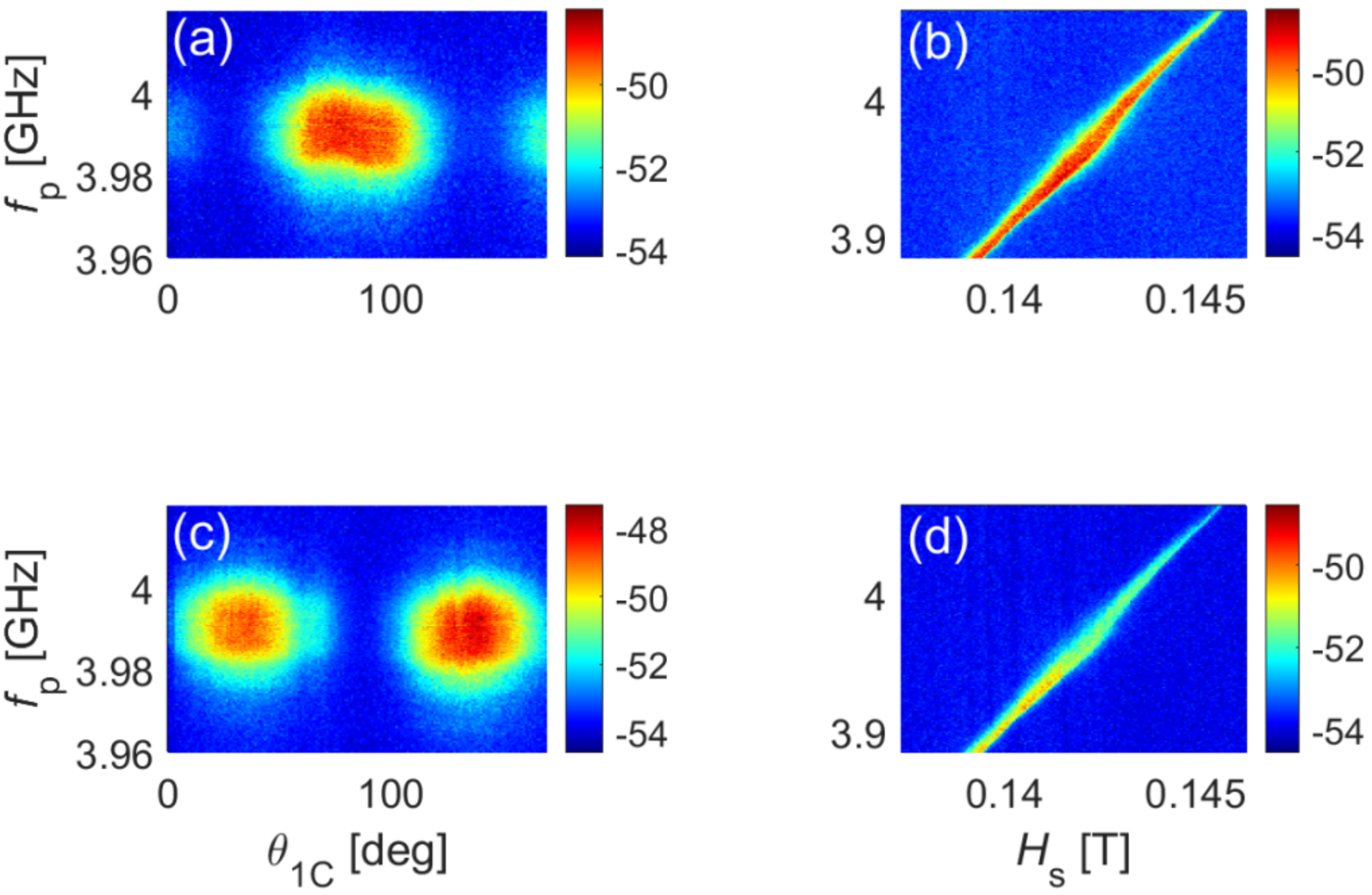}
\end{center}
\caption{Side bands in dBm units. (a) anti-Stokes intensity as a function of
PC1 angle $\theta_{1\mathrm{C}}$. (b) anti-Stokes intensity as a function of
magnetic field $H_{\mathrm{s}}$. (c) Stokes intensity as a function of
$\theta_{1\mathrm{C}}$. (d) Stokes intensity as a function of $H_{\mathrm{s}}%
$. The magnetic field $H_{\mathrm{s}}$ in (a) and (c) is tuned near
avoided-crossing regime. TL1 is set at optical power of $31\operatorname{mW}$
and wavelength of $\lambda_{\mathrm{L}}$ of $1537.7\operatorname{nm}$, and the
driving microwave power is set at $P_{\mathrm{p}}=20$ dBm. In (a) and (c),
$\theta_{1\mathrm{A}}=170^{\circ}$ and $\theta_{1\mathrm{B}}=85^{\circ}$, and
$\theta_{1\mathrm{C}}$ is varied from $0^{\circ}$ to $170^{\circ}$, whereas in
(b) and (d) $\left(  \theta_{1\mathrm{A}},\theta_{1\mathrm{B}},\theta
_{1\mathrm{C}}\right)  =\left(  170^{\circ},85^{\circ},60^{\circ}\right)  $
(for this setting both Stokes and anti-Stokes peaks are clearly visible near
the FSR resonance).}%
\label{Figsubplots}%
\end{figure}

Both motorized polarization controllers (labelled as PC1 and PC2 in
Fig.~\ref{FigExSetup}) have three optomechanical components (paddles), which
act as either quarter or half wave plates. The paddles' angles of PC1 (PC2)
are denoted by $\theta_{1\mathrm{A}}$, $\theta_{1\mathrm{B}}$ and
$\theta_{1\mathrm{C}}$ ($\theta_{2\mathrm{A}}$, $\theta_{2\mathrm{B}}$ and
$\theta_{2\mathrm{C}}$). The incident light state of polarization (SOP) can be
manipulated using PC1. We observe that intensity of lower wavelength
$\lambda_{\mathrm{L}}-\lambda_{\mathrm{SB}}$ anti-Stokes sideband and higher
wavelength $\lambda_{\mathrm{L}}+\lambda_{\mathrm{SB}}$ Stokes side band
depend on the input SOP. SSM in the transmission spectrum, with either single
anti-Stokes side band, or with single Stokes side band, can be obtained by
adjusting PC1. The plot shown in Fig.~\ref{Figsubplots}(a) exhibits the
measured anti-Stokes side band intensity as a function of microwave driving
frequency $f_{\mathrm{p}}=\omega_{\mathrm{p}}/\left(  2\pi\right)  $ and PC1
angle $\theta_{1\mathrm{C}}$ near the avoided-crossing region. The plot shown
in Fig.~\ref{Figsubplots}(c) exhibits simultaneously measured Stokes side band
intensity in the same region. We clearly observe appreciable anti-Stokes and
Stokes intensity in Fig.~\ref{Figsubplots} (a) and (c), respectively, when
driving frequency $\omega_{\mathrm{p}}/\left(  2\pi\right)  $ becomes close to
FSR resonance $\omega_{\mathrm{m}}/\left(  2\pi\right)  $. However, they are
asymmetric. For a certain range of PC1 position, SSM is obtained, i.e. only
one side band, either anti-Stokes or Stokes, is observed. Contrary to other
experimental setups, in which the FSR is optically coupled by either a tapered
optical fiber or a prism, for our setup, for which the measured optical
transmission only weakly depends on the input wavelength $\lambda_{\mathrm{L}%
}$, SSM can be obtained in wide range of $\lambda_{\mathrm{L}}$.

A rotating lambda plate polarimeter is employed to measure the input SOP. The
polarimeter measurements reveal that the input SOP for the two extreme cases
(SSM of either anti-Stokes or Stokes peak) are orthogonal to each other (i.e.
separated by a diameter on the Poincar\'{e} sphere).

The plots shown in Fig.~\ref{Figsubplots} (b) and (d) exhibit anti-Stokes and
Stokes intensity, respectively, as a function of microwave driving frequency
$f_{\mathrm{p}}$ and static magnetic field $H_{\mathrm{s}}$. The FSR resonance
changes as we vary the static magnetic field $H_{\mathrm{s}}$. Accordingly,
from Fig.~\ref{Figsubplots}(b) and (d), we see that both anti-Stokes and
Stokes intensity gets pronounced when driving frequency $\omega_{\mathrm{p}%
}/\left(  2\pi\right)  $ is within the bandwidth of FSR resonance at
$\omega_{\mathrm{m}}/\left(  2\pi\right)  $.

\begin{figure}[ptb]
\begin{center}
\includegraphics[width=3.2in,keepaspectratio]{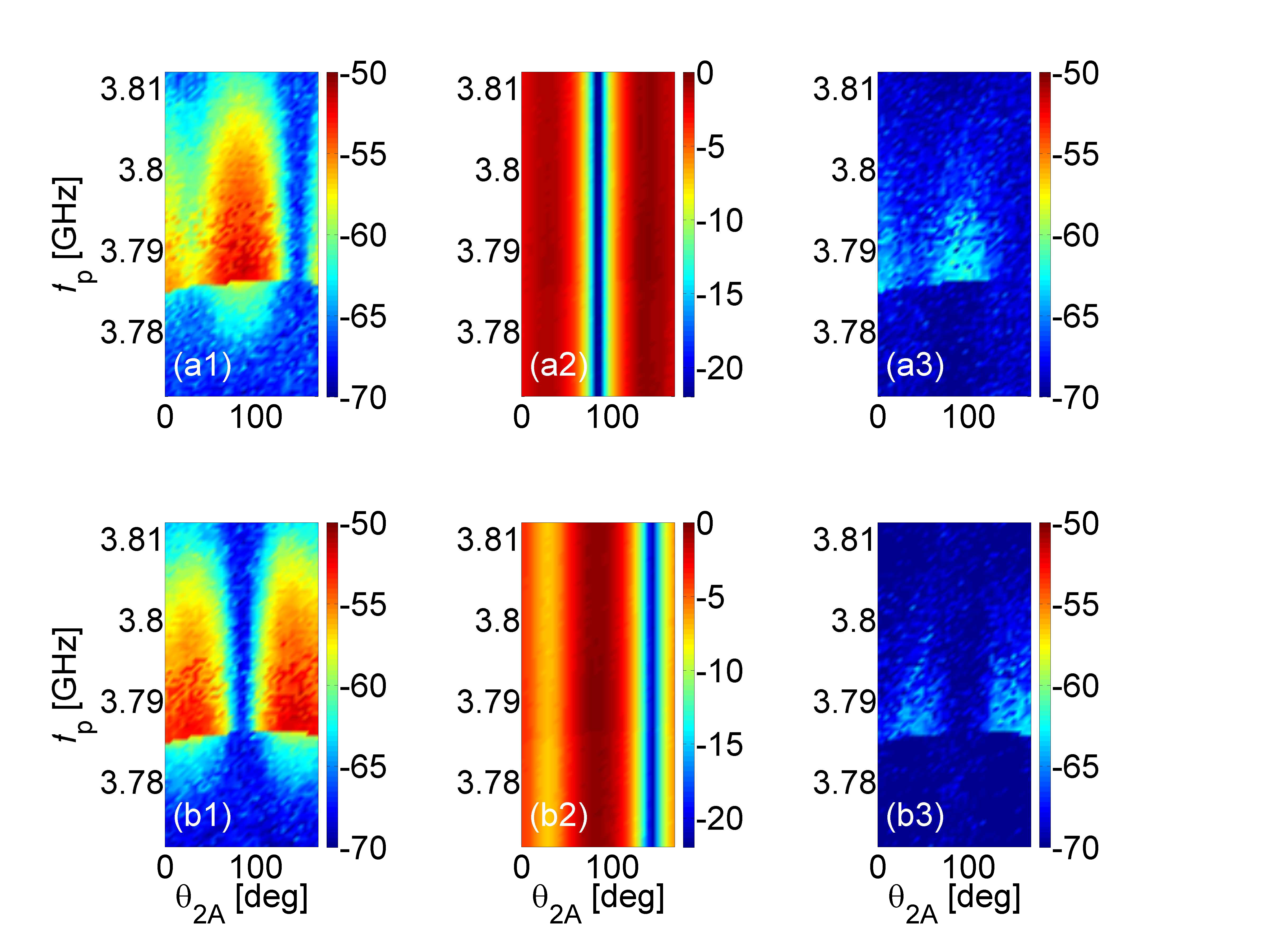}
\end{center}
\caption{Sideband SOP. The measured intensity $I_{\mathrm{DPD1}}$
($I_{\mathrm{DPD2}}$) is shown (in dBm units) in the plots labeled by the
letter 'a' ('b'). The intensity at wavelengths $\lambda_{\mathrm{L}}%
-\lambda_{\mathrm{SB}}$, $\lambda_{\mathrm{L}}$ and $\lambda_{\mathrm{L}%
}+\lambda_{\mathrm{SB}}$ is shown in the plots labelled by the numbers '1',
'2' and '3', respectively. The TL1 is set at optical power of
$31\operatorname{mW}$ and wavelength of $\lambda_{\mathrm{L}}$ of
$1538.9\operatorname{nm}$, the driving microwave is set at power
$P_{\mathrm{p}}$ of $20$ dBm.}%
\label{FigPC6}%
\end{figure}

Our experimental setup (see Fig.~\ref{FigExSetup}) allows measuring the SOP of
both sidebands. While the plots shown in Fig.~\ref{Figsubplots} display the
total optical intensity $I_{\mathrm{T}}=I_{\mathrm{DPD1}}+I_{\mathrm{DPD2}}$,
the intensity $I_{\mathrm{DPD1}}$ ($I_{\mathrm{DPD2}}$) is separately
displayed in the top (bottom) row plots shown in Fig.~\ref{FigPC6}. These two
intensities $I_{\mathrm{DPD1}}$ and $I_{\mathrm{DPD2}}$ represent two
orthogonal SOP, which can be set by adjusting PC2 (see Fig.~\ref{FigExSetup}).
The left (right) column plots in Fig.~\ref{FigPC6} display the measured
intensity of the left anti-Stokes (right Stokes) sideband at wavelength
$\lambda_{\mathrm{L}}-\lambda_{\mathrm{SB}}$ ($\lambda_{\mathrm{L}}%
+\lambda_{\mathrm{SB}}$), whereas the intensity at the central wavelength
$\lambda_{\mathrm{L}}$ is displayed by the central column plots in
Fig.~\ref{FigPC6}. For the measurements shown in Fig.~\ref{FigPC6}, PC1 is set
to a nearly SSM state. By varying the setting of PC2, we find that the central
peak at wavelength $\lambda_{\mathrm{L}}$ is maximized (minimized) in the same
region where the sidebands at wavelength $\lambda_{\mathrm{L}}\pm
\lambda_{\mathrm{SB}}$ are minimized (maximized). This observation implies
that in the region of SSM, the SOP of the sidebands is nearly orthogonal to
the SOP of the incident light. This orthogonality can be exploited at the
receiver end of a data transmission system based on our proposed MO
modulation, since it allows demodulation by polarization filtering-out of the
carrier at wavelength $\lambda_{\mathrm{L}}$.

\section{MO coupling}

The MO coupling giving rise to the optical sidebands originates from an
interaction term in the system's Hamiltonian, which is denoted by
$V_{\mathrm{SB}}$. This term $V_{\mathrm{SB}}$ is commonly derived from the
classical energy density associated with the interaction between magnetization
and optical modes. For the case where only whispering gallery FSR optical
modes participate in the interaction, the term $V_{\mathrm{SB}}$ was derived
in
\cite{Haigh_133602,Osada_103018,Osada_223601,Sharma_094412,Almpanis_184406,Zivieri_165406,Desormiere_379,Liu_3698,Chai_820}%
, whereas for our experimental configuration we consider the case where light
propagates through the FSR bulk.

Consider an incident $\mathrm{I}$ (scattered $\mathrm{S}$) optical field,
having right and left handed circular polarization amplitudes $E_{\mathrm{I}%
+}^{{}}$ and $E_{\mathrm{I}-}^{{}}$ ($E_{\mathrm{S}+}^{{}}$ and $E_{\mathrm{S}%
-}^{{}}$), respectively. The time-averaged energy density $u_{\mathrm{m}}$
associated with MO coupling is given by $u_{\mathrm{m}}=\left(  1/4\right)
\operatorname{Re}\mathcal{U}_{\mathrm{m}}$, where%
\begin{equation}
\mathcal{U}_{\mathrm{m}}=\left(
\begin{array}
[c]{cc}%
E_{\mathrm{S}+}^{\ast} & E_{\mathrm{S}-}^{\ast}%
\end{array}
\right)  \epsilon_{\mathrm{m}}\left(
\begin{array}
[c]{c}%
E_{\mathrm{I}+}^{{}}\\
E_{\mathrm{I}-}^{{}}%
\end{array}
\right)  \;, \label{Um}%
\end{equation}
and where $\epsilon_{\mathrm{m}}=\epsilon_{\mathrm{m0}}+\epsilon_{\mathrm{m+}%
}m_{+^{\prime}}+\epsilon_{\mathrm{m-}}m_{-^{\prime}}$ is a transverse
permittivity tensor. The static part $\epsilon_{\mathrm{m0}}$ is given by
Eq.~(\ref{ep_mo}) of appendix \ref{AppTPT}. The diagonal elements of
$\epsilon_{\mathrm{m0}}$ give rise to the static Faraday effect, whereas the
static Voigt (Cotton-Mouton) effect originates from the off-diagonal elements
of $\epsilon_{\mathrm{m0}}$ [see Eq.~(\ref{ep_mo})]. The terms $\epsilon
_{\mathrm{m+}}m_{+^{\prime}}$ and $\epsilon_{\mathrm{m-}}m_{-^{\prime}}$
account for the effect of magnetization precession, where $\epsilon
_{\mathrm{m\pm}}$ is given by Eq.~(\ref{ep_m pm}) of appendix \ref{AppTPT},
and $m_{\pm^{\prime}}$ represent amplitudes of magnetization precession. Note
that the matrix $\epsilon_{\mathrm{m\pm}}$ is proportional to $e^{\pm
i\varphi}$, where $\varphi$ is the azimuthal angle [see Eq.~(\ref{ep_m pm})].
The spherical symmetry of the FSR is partially broken by the two CFs that are
employed for holding it (see Fig.~\ref{FigExSetup}).

In the semiclassical approximation $V_{\mathrm{SB}}$ is derived from
$u_{\mathrm{m}}=\left(  1/4\right)  \operatorname{Re}\mathcal{U}_{\mathrm{m}}$
[see Eq.~(\ref{Um})]. Consider a pair of optical modes having normalized
scalar spatial waveforms, which in spherical coordinates are expressed as
$u_{n^{\prime}}\left(  r,\theta,\varphi\right)  $ and $u_{n^{\prime\prime}%
}\left(  r,\theta,\varphi\right)  $, respectively. The contribution of this
pair to the total interaction term $V_{\mathrm{SB}}$, which is denoted by
$V_{n^{\prime},n^{\prime\prime}}$, is expressed as
\begin{equation}
V_{n^{\prime},n^{\prime\prime}}=a_{n^{\prime}}^{\dag}a_{n^{\prime\prime}}^{{}%
}\left(  g_{n^{\prime},n^{\prime\prime},+}b^{\dag}+g_{n^{\prime}%
,n^{\prime\prime},-}b^{{}}\right)  +\mathrm{h.c.}\;,
\end{equation}
where $a_{n}$ ($b$) is an annihilation operator for the $n$'th optical mode
(magnon mode), and $\mathrm{h.c.}$ stands for Hermitian conjugate. The
coupling coefficients $g_{n^{\prime},n^{\prime\prime},\pm}$ are given by
(recall that in our experiment the static magnetic field is normal to the
light propagation direction)%
\begin{equation}
\hbar^{-1}g_{n^{\prime},n^{\prime\prime},\pm}\simeq g_{0}\int\mathrm{d}%
\mathbf{r}^{\prime}\;e^{\pm i\varphi}u_{n^{\prime}}^{{}}\left(  \mathbf{r}%
^{\prime}\right)  u_{n^{\prime\prime}}^{\ast}\left(  \mathbf{r}^{\prime
}\right)  \;, \label{g n',n''}%
\end{equation}
where $g_{0}=\omega_{\mathrm{L}}Q_{\mathrm{s}}/\left(  8n_{0}^{2}%
N_{\mathrm{s}}^{1/2}\right)  $, and $N_{\mathrm{s}}$ is the number of FSR
spins ($N_{\mathrm{s}}=3.4\times10^{16}$ for the FSR under study). For YIG in
the telecom band (free space wavelength $\lambda_{0}\simeq1550%
\operatorname{nm}%
$), the refractive index is $n_{0}=2.19$, and the dimensionless MO coupling
coefficient is $Q_{\mathrm{s}}\simeq10^{-4}$ \cite{Wood_1038}, and thus
$g_{0}/\left(  2\pi\right)  =2.7%
\operatorname{Hz}%
$. The overlap integral in Eq.~(\ref{g n',n''}) represents a photon-magnon
scattering selection rule \cite{Haigh_133602,Osada_103018}.

The ratio of side band output optical power to the input optical power is
denoted by $\eta_{\mathrm{SB}}$. The largest value of $\eta_{\mathrm{SB}}$ is
obtained at the triple resonance \cite{Haigh_133602}, for which the MW driving
is tuned to the FSR resonance $\omega_{\mathrm{m}}$, the laser frequency
$\omega_{\mathrm{L}}$ matches the frequency of one optical mode, and the
second one has a frequency detuned from $\omega_{\mathrm{L}}$ by
$\omega_{\mathrm{m}}$. For this case $\eta_{\mathrm{SB}}\simeq\left(
2n_{0}R_{\mathrm{s}}g_{0}/c\right)  ^{2}N_{\mathrm{m}}$ [it is assumed that
the overlap integral in Eq.~(\ref{g n',n''}) is of order unity], where
$N_{\mathrm{m}}$ is the averaged number of excited magnons in steady state.
For the case where the MWA is nearly critically coupled to the FSR, at
resonance $N_{\mathrm{m}}\simeq P_{\mathrm{p}}/\left(  \hbar\omega
_{\mathrm{m}}\kappa_{\mathrm{m}}\right)  $, where $\kappa_{\mathrm{m}}$ is the
FSR damping rate. The values of $P_{\mathrm{p}}=20$ dBm, $\omega_{\mathrm{m}%
}/\left(  2\pi\right)  =3.8%
\operatorname{GHz}%
$ and $\kappa_{\mathrm{m}}/\left(  2\pi\right)  =1%
\operatorname{MHz}%
$ yield $\eta_{\mathrm{SB}}\simeq10^{-5}$. This rough estimate agrees with the
experimentally observed value of $\eta_{\mathrm{SB}}$ [see Fig.~\ref{FigPC6}].

\section{Kerr nonlinearity}

\begin{figure}[ptb]
\begin{center}
\includegraphics[width=3.2in,keepaspectratio]{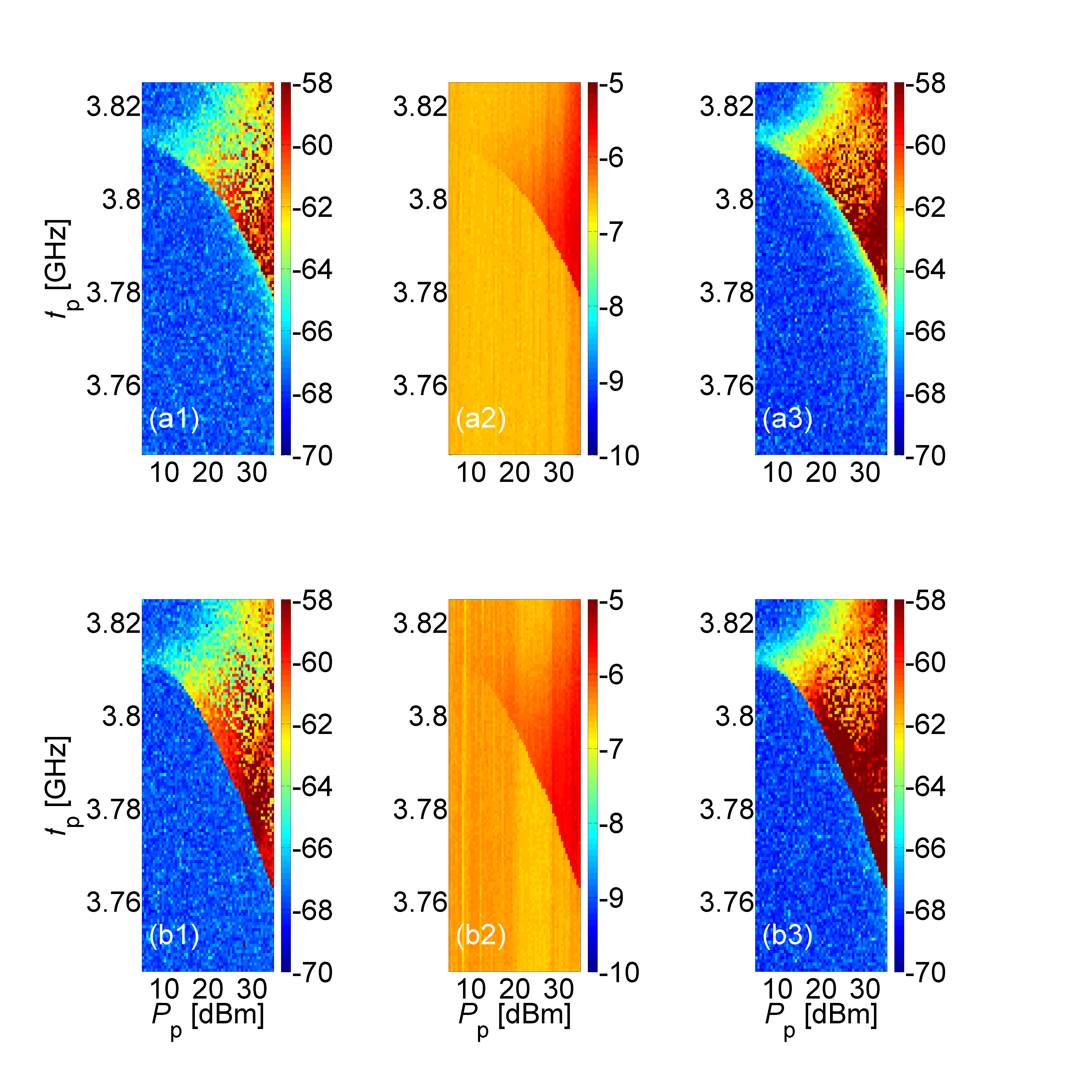}
\end{center}
\caption{{}Spectral peaks (in dBm units) in the nonlinear regime as a function
of MW driving power $P_{\mathrm{p}}$. The intensity of the left (right)
sideband at wavelength $\lambda_{\mathrm{L}}-\lambda_{\mathrm{SB}}$
($\lambda_{\mathrm{L}}+\lambda_{\mathrm{SB}}$) is shown in the plots in the
left (right) column, whereas the plots in the central column show the
intensity of the central optical peak (at TL1 wavelength $\lambda_{\mathrm{L}%
}$). For the plots shown in the top (bottom) row, the frequency $f_{\mathrm{p}%
}$ is swept upwards (downwards).}%
\label{FigNL}%
\end{figure}

Magnetic anisotropy gives rise to Kerr nonlinearity in the FSR response
\cite{Wang_224410}. The nonlinearity can be exploited for modulation
amplification \cite{Mathai_67001}. Modulation measurements in the nonlinear
regime are shown in Fig.~\ref{FigNL}. The results indicate that the Kerr
coefficient is negative (giving rise to softening). For the plots shown in the
top (bottom) row of Fig.~\ref{FigNL}, the microwave driving frequency is swept
upwards (downwards). The dependency on sweeping direction is attributed to
nonlinearity-induced bistability, which, in turn, gives rise to hysteresis.

\section{Summary}

In summary, polarization-selective SSM in the telecom band is achieved using
an FSR strongly coupled to an LGR. The modulator can be used in a wide optical band, and it is compatible with ultra low temperatures. Future study will explore potential applications, including quantum state readout of superconducting circuits.

This work was supported by the Israeli science foundation, the Israeli
ministry of science, and by the Technion security research foundation.

\appendix

\section{Transverse permittivity tensor}

\label{AppTPT}

The evolution of electromagnetic waves propagating inside a magnetized medium
is governed by a $3\times3$ permittivity tensor
\cite{Freiser_152,Boardman_197,Boardman_388}. Consider a Cartesian coordinate
system $\left(  x,y,z\right)  $, for which the propagation direction is
parallel to the $z$ direction. In this system the static magnetic field
(magnetization vector) is parallel to a unit vector denoted by $\mathbf{\hat
{h}}$ ($\mathbf{\hat{m}}$). The angle between $\mathbf{\hat{h}}=\left(
h_{x},h_{y},h_{z}\right)  =\left(  \sin\theta\cos\varphi,\sin\theta\sin
\varphi,\cos\theta\right)  $\ and $\mathbf{\hat{m}}=\left(  m_{x},m_{y}%
,m_{z}\right)  $\ is assumed to be small.

From the $3\times3$ permittivity tensor, a $2\times2$ transverse permittivity
tensor $\epsilon_{\mathrm{T}}$ can be derived. In a basis of circular SOP
$\epsilon_{\mathrm{T}}$ is given by $\epsilon_{\mathrm{T}}=n_{0}^{2}%
I+\epsilon_{\mathrm{m}}$, where $n_{0}$ is the medium refractive index, $I$ is
the $2\times2$ identity matrix, and the $2\times2$ matrix $\epsilon
_{\mathrm{m}}$ (in a basis of circular SOPs) is given by \cite{Buks_014421}%
\begin{equation}
\frac{\epsilon_{\mathrm{m}}}{n_{0}^{2}}=\left(
\begin{array}
[c]{cc}%
Q_{\mathrm{s}}m_{z} & Q_{\mathrm{s}}^{2}m_{-}^{2}\\
Q_{\mathrm{s}}^{2}m_{+}^{2} & -Q_{\mathrm{s}}m_{z}%
\end{array}
\right)  \;, \label{ep_m}%
\end{equation}
where $m_{\pm}=\left(  m_{x}\pm im_{y}\right)  /\sqrt{2}$. For YIG in the
telecom band, the refractive index is $n_{0}=2.19$, and the dimensionless MO
coupling coefficient is $Q_{\mathrm{s}}\simeq10^{-4}$ \cite{Wood_1038}.

The eigenvalues of $\epsilon_{\mathrm{m}}/n_{0}^{2}$ (\ref{ep_m}) are given by
$\pm Q_{\mathrm{s}}\sqrt{m_{z}^{2}+Q_{\mathrm{s}}^{2}m_{-}^{2}m_{+}^{2}}$. For
the Faraday configuration, for which $m_{x}=m_{y}=0$ and $m_{z}=1$, i.e.
$\mathbf{\hat{m}}$ is parallel to the propagation direction, the eigenvectors
of $\epsilon_{\mathrm{m}}/n_{0}^{2}$ represent circular SOPs, the
corresponding eigenvalues are $\pm Q_{\mathrm{s}}$, and MO coupling gives rise
to circular birefringence, whereas for the Voigt (Cotton-Mouton)
configuration, for which $m_{z}=0$ and $m_{x}^{2}+m_{y}^{2}=1$, i.e.
$\mathbf{\hat{m}}$ is perpendicular to the propagation direction, the
eigenvectors of $\epsilon_{\mathrm{m}}/n_{0}^{2}$ represent colinear SOPs, the
corresponding eigenvalues are $\pm Q_{\mathrm{s}}^{2}/2$ [note that $m_{-}%
^{2}m_{+}^{2}=\left(  m_{x}^{2}+m_{y}^{2}\right)  ^{2}/4$], and MO coupling
gives rise to colinear birefringence. Note that for the Faraday configuration,
the SOP rotation angle that is accumulated over a traveling distance of a
single optical wavelength is $2\pi Q_{\mathrm{s}}$.

To describe the effect of magnetization precession on $\epsilon_{\mathrm{m}}$,
it is convenient to express $\mathbf{\hat{m}}$ (magnetization unit vector) as
a sum of parallel and perpendicular components, with respect to $\mathbf{\hat
{h}}$ (magnetic field unit vector). In a Cartesian coordinate system $\left(
x^{\prime},y^{\prime},z^{\prime}\right)  $, for which the static magnetic
field is parallel to the $z^{\prime}$ direction, the unit vector parallel to
the magnetization vector is expressed as $\mathbf{\hat{m}}^{\prime
}=m_{x^{\prime}}\mathbf{\hat{x}}^{\prime}+m_{y^{\prime}}\mathbf{\hat{y}%
}^{\prime}+m_{z^{\prime}}\mathbf{\hat{z}}^{\prime}=m_{+^{\prime}}%
\mathbf{\hat{u}}_{+}^{\prime}+m_{-^{\prime}}\mathbf{\hat{u}}_{-}^{\prime
}+m_{z^{\prime}}\mathbf{\hat{z}}^{\prime}$, where $\mathbf{\hat{u}}_{\pm
}^{\prime}=\left(  \mathbf{\hat{x}}^{\prime}\pm i\mathbf{\hat{y}}^{\prime
}\right)  /\sqrt{2}$, and where $m_{\pm^{\prime}}=\left(  m_{x^{\prime}}\mp
im_{y^{\prime}}\right)  /\sqrt{2}$. The unit vectors $\mathbf{\hat{m}}$ and
$\mathbf{\hat{m}}^{\prime}$\ are related by $\mathbf{\hat{m}}=R_{\mathbf{\hat
{h}}}^{-1}\mathbf{\hat{m}}^{\prime}$, where for a given unit vector
$\mathbf{\hat{n}}$, the rotation matrix $R_{\mathbf{\hat{n}}}$ is defined by
the relation $R_{\mathbf{\hat{n}}}\mathbf{\hat{n}}=\mathbf{\hat{z}}$, and thus
$\mathbf{\hat{m}}=m_{+^{\prime}}\mathbf{\hat{v}}_{+}+m_{-^{\prime}%
}\mathbf{\hat{v}}_{-}+m_{z^{\prime}}R_{\mathbf{\hat{h}}}^{-1}\mathbf{\hat{z}%
}^{\prime}$, where $\mathbf{\hat{v}}_{\pm}=R_{\mathbf{\hat{h}}}^{-1}%
\mathbf{\hat{u}}_{\pm}^{\prime}$. The matrix elements of the $3\times3$
rotation matrix $R_{\mathbf{\hat{h}}}$ are given by $R_{11}=1+\left(
\cos\theta-1\right)  \cos^{2}\varphi$, $R_{22}=1+\left(  \cos\theta-1\right)
\sin^{2}\varphi$, $R_{12}=R_{21}=\left(  1/2\right)  \left(  \cos
\theta-1\right)  \sin\left(  2\varphi\right)  $, $R_{31}=-R_{13}=\sin
\theta\cos\varphi$, $R_{32}=-R_{23}=\sin\theta\sin\varphi$ and $R_{33}%
=\cos\theta$. The following holds $\mathbf{\hat{v}}_{\pm}=\cos^{2}\left(
\theta/2\right)  \mathbf{\hat{u}}_{\pm}-e^{\pm2i\varphi}\sin^{2}\left(
\theta/2\right)  \mathbf{\hat{u}}_{\mp}-2^{-1/2}e^{\pm i\varphi}\left(
\sin\theta\right)  \mathbf{\hat{z}}$, hence $\mathbf{\hat{m}}=\mu
_{+}\mathbf{\hat{u}}_{+}+\mu_{-}\mathbf{\hat{u}}_{-}+\mu_{z}\mathbf{\hat{z}%
}+m_{z^{\prime}}\mathbf{\hat{h}}$, where $\mu_{\pm}=m_{\pm^{\prime}}\cos
^{2}\left(  \theta/2\right)  -m_{\mp^{\prime}}e^{\mp2i\varphi}\sin^{2}\left(
\theta/2\right)  $ and $\mu_{z}=-2^{-1/2}\left(  m_{+^{\prime}}e^{i\varphi
}+m_{-^{\prime}}e^{-i\varphi}\right)  \sin\theta$, and thus $m_{\pm}=\mu_{\mp
}+2^{-1/2}m_{z^{\prime}}e^{\mp i\varphi}\sin\theta$.

The assumption that the angle between the static magnetic field\ and the
magnetization vector\ is small implies that $m_{z^{\prime}}\simeq1$ and
$\left\vert m_{\pm^{\prime}}\right\vert \ll1$. To first order in $\left\vert
m_{\pm^{\prime}}\right\vert $, $\epsilon_{\mathrm{m}}$ can be expanded as
$\epsilon_{\mathrm{m}}=\epsilon_{\mathrm{m0}}+\epsilon_{\mathrm{m+}%
}m_{+^{\prime}}+\epsilon_{\mathrm{m-}}m_{-^{\prime}}$, where $\epsilon
_{\mathrm{m0}}$, which is given by [compare with Eq.~(\ref{ep_m})]%
\begin{equation}
\frac{\epsilon_{\mathrm{m0}}}{n_{0}^{2}}=\left(
\begin{array}
[c]{cc}%
Q_{\mathrm{s}}\cos\theta & \frac{Q_{\mathrm{s}}^{2}e^{2i\varphi}\sin^{2}%
\theta}{2}\\
\frac{Q_{\mathrm{s}}^{2}e^{-2i\varphi}\sin^{2}\theta}{2} & -Q_{\mathrm{s}}%
\cos\theta
\end{array}
\right)  \;,\label{ep_mo}%
\end{equation}
accounts for static magnetization, and where $\epsilon_{\mathrm{m\pm}}$, which is given by%
\begin{equation}
\frac{\epsilon_{\mathrm{m\pm}}}{n_{0}^{2}}=\frac{Q_{\mathrm{s}}e^{\pm
i\varphi}\sin\theta}{\sqrt{2}}\left(
\begin{array}
[c]{cc}%
-1 & \pm Q_{\mathrm{s}}\left(  1\pm\cos\theta\right)  \\
\mp Q_{\mathrm{s}}\left(  1\mp\cos\theta\right)   & 1
\end{array}
\right)  \;,\label{ep_m pm}%
\end{equation}
accounts for magnetization precession.

\bibliographystyle{ieeepes}
\bibliography{acompat,Eyal_Bib}

\newif\ifabfull\abfulltrue
\begin{thebibliography}{10}

\bibitem{lazzarini2008asymmetric}
Victor Lazzarini, Joseph Timoney, and Thomas Lysaght,
\newblock ``Asymmetric-spectra methods for adaptive fm synthesis'',
\newblock 2008.

\bibitem{Li_1}
Wei Li, Wen~Ting Wang, Li~Xian Wang, and Ning~Hua Zhu,
\newblock ``Optical vector network analyzer based on single-sideband modulation
  and segmental measurement'',
\newblock {\em IEEE Photonics Journal}, vol. 6, no. 2, pp. 1--8, 2014.

\bibitem{Shimotsu_364}
S~Shimotsu, S~Oikawa, T~Saitou, N~Mitsugi, K~Kubodera, T~Kawanishi, and
  M~Izutsu,
\newblock ``Single side-band modulation performance of a linbo 3 integrated
  modulator consisting of four-phase modulator waveguides'',
\newblock {\em IEEE Photonics Technology Letters}, vol. 13, no. 4, pp.
  364--366, 2001.

\bibitem{Rameshti_1}
Babak~Zare Rameshti, Silvia~Viola Kusminskiy, James~A Haigh, Koji Usami, Dany
  Lachance-Quirion, Yasunobu Nakamura, Can-Ming Hu, Hong~X Tang, Gerrit~EW
  Bauer, and Yaroslav~M Blanter,
\newblock ``Cavity magnonics'',
\newblock {\em Physics Reports}, vol. 979, pp. 1--61, 2022.

\bibitem{Kusminskiy_299}
Silvia~Viola Kusminskiy,
\newblock ``Cavity optomagnonics'',
\newblock in {\em Optomagnonic Structures: Novel Architectures for Simultaneous
  Control of Light and Spin Waves}, pp. 299--353. World Scientific, 2021.

\bibitem{Zhu_2012_11119}
Na~Zhu, Xufeng Zhang, Xu~Han, Chang-Ling Zou, and Hong~X Tang,
\newblock ``Inverse faraday effect in an optomagnonic waveguide'',
\newblock {\em arXiv:2012.11119}, 2020.

\bibitem{Juraschek_094407}
Dominik~M Juraschek, Derek~S Wang, and Prineha Narang,
\newblock ``Sum-frequency excitation of coherent magnons'',
\newblock {\em Physical Review B}, vol. 103, no. 9, pp. 094407, 2021.

\bibitem{Bittencourt_014409}
VASV Bittencourt, I~Liberal, and S~Viola Kusminskiy,
\newblock ``Light propagation and magnon-photon coupling in optically
  dispersive magnetic media'',
\newblock {\em Physical Review B}, vol. 105, no. 1, pp. 014409, 2022.

\bibitem{Zhang_123605}
Xufeng Zhang, Na~Zhu, Chang-Ling Zou, and Hong~X Tang,
\newblock ``Optomagnonic whispering gallery microresonators'',
\newblock {\em Physical review letters}, vol. 117, no. 12, pp. 123605, 2016.

\bibitem{Stancil_Spin}
Daniel~D Stancil and Anil Prabhakar,
\newblock {\em Spin waves},
\newblock Springer, 2009.

\bibitem{Haigh_133602}
JA~Haigh, Andreas Nunnenkamp, AJ~Ramsay, and AJ~Ferguson,
\newblock ``Triple-resonant brillouin light scattering in magneto-optical
  cavities'',
\newblock {\em Physical review letters}, vol. 117, no. 13, pp. 133602, 2016.

\bibitem{Osada_103018}
A~Osada, A~Gloppe, Y~Nakamura, and K~Usami,
\newblock ``Orbital angular momentum conservation in brillouin light scattering
  within a ferromagnetic sphere'',
\newblock {\em New Journal of Physics}, vol. 20, no. 10, pp. 103018, 2018.

\bibitem{Osada_223601}
A~Osada, R~Hisatomi, A~Noguchi, Y~Tabuchi, R~Yamazaki, K~Usami, M~Sadgrove,
  R~Yalla, M~Nomura, and Y~Nakamura,
\newblock ``Cavity optomagnonics with spin-orbit coupled photons'',
\newblock {\em Physical review letters}, vol. 116, no. 22, pp. 223601, 2016.

\bibitem{Sharma_094412}
Sanchar Sharma, Yaroslav~M Blanter, and Gerrit~EW Bauer,
\newblock ``Light scattering by magnons in whispering gallery mode cavities'',
\newblock {\em Physical Review B}, vol. 96, no. 9, pp. 094412, 2017.

\bibitem{Almpanis_184406}
Evangelos Almpanis,
\newblock ``Dielectric magnetic microparticles as photomagnonic cavities:
  Enhancing the modulation of near-infrared light by spin waves'',
\newblock {\em Physical Review B}, vol. 97, no. 18, pp. 184406, 2018.

\bibitem{Zivieri_165406}
R~Zivieri, P~Vavassori, L~Giovannini, F~Nizzoli, Eric~E Fullerton,
  M~Grimsditch, and V~Metlushko,
\newblock ``Stokes--anti-stokes brillouin intensity asymmetry of spin-wave
  modes in ferromagnetic films and multilayers'',
\newblock {\em Physical Review B}, vol. 65, no. 16, pp. 165406, 2002.

\bibitem{Desormiere_379}
BERNARD Desormiere and HENRI Le~Gall,
\newblock ``Interaction studies of a laser light with spin waves and
  magnetoelastic waves propagating in a yig bar'',
\newblock {\em IEEE Transactions on Magnetics}, vol. 8, no. 3, pp. 379--381,
  1972.

\bibitem{Liu_3698}
Zeng-Xing Liu, Bao Wang, Hao Xiong, and Ying Wu,
\newblock ``Magnon-induced high-order sideband generation'',
\newblock {\em Optics Letters}, vol. 43, no. 15, pp. 3698--3701, 2018.

\bibitem{Chai_820}
Cheng-Zhe Chai, Zhen Shen, Yan-Lei Zhang, Hao-Qi Zhao, Guang-Can Guo,
  Chang-Ling Zou, and Chun-Hua Dong,
\newblock ``Single-sideband microwave-to-optical conversion in high-q
  ferrimagnetic microspheres'',
\newblock {\em Photonics Research}, vol. 10, no. 3, pp. 820--827, 2022.

\bibitem{Zhu_1291}
Na~Zhu, Xufeng Zhang, Xu~Han, Chang-Ling Zou, Changchun Zhong, Chiao-Hsuan
  Wang, Liang Jiang, and Hong~X Tang,
\newblock ``Waveguide cavity optomagnonics for microwave-to-optics
  conversion'',
\newblock {\em Optica}, vol. 7, no. 10, pp. 1291--1297, 2020.

\bibitem{Li_040344}
Jie Li, Yi-Pu Wang, Wei-Jiang Wu, Shi-Yao Zhu, and JQ~You,
\newblock ``Quantum network with magnonic and mechanical nodes'',
\newblock {\em PRX Quantum}, vol. 2, no. 4, pp. 040344, 2021.

\bibitem{Wettling_211}
W~Wettling, MG~Cottam, and JR~Sandercock,
\newblock ``The relation between one-magnon light scattering and the complex
  magneto-optic effects in yig'',
\newblock {\em Journal of Physics C: Solid State Physics}, vol. 8, no. 2, pp.
  211, 1975.

\bibitem{cottam1986light}
Michael~G Cottam and David~J Lockwood,
\newblock {\em Light scattering in magnetic solids},
\newblock Wiley New York, 1986.

\bibitem{Liu_060405}
Tianyu Liu, Xufeng Zhang, Hong~X Tang, and Michael~E Flatt{\'e},
\newblock ``Optomagnonics in magnetic solids'',
\newblock {\em Physical Review B}, vol. 94, no. 6, pp. 060405, 2016.

\bibitem{Haigh_143601}
JA~Haigh, A~Nunnenkamp, and AJ~Ramsay,
\newblock ``Polarization dependent scattering in cavity optomagnonics'',
\newblock {\em Physical Review Letters}, vol. 127, no. 14, pp. 143601, 2021.

\bibitem{Hisatomi_207401}
R~Hisatomi, A~Noguchi, R~Yamazaki, Y~Nakata, A~Gloppe, Y~Nakamura, and K~Usami,
\newblock ``Helicity-changing brillouin light scattering by magnons in a
  ferromagnetic crystal'',
\newblock {\em Physical Review Letters}, vol. 123, no. 20, pp. 207401, 2019.

\bibitem{Hisatomi_174427}
Ryusuke Hisatomi, Alto Osada, Yutaka Tabuchi, Toyofumi Ishikawa, Atsushi
  Noguchi, Rekishu Yamazaki, Koji Usami, and Yasunobu Nakamura,
\newblock ``Bidirectional conversion between microwave and light via
  ferromagnetic magnons'',
\newblock {\em Physical Review B}, vol. 93, no. 17, pp. 174427, 2016.

\bibitem{Baney_355}
Douglas~M Baney, Bogdan Szafraniec, and Ali Motamedi,
\newblock ``Coherent optical spectrum analyzer'',
\newblock {\em Ieee Photonics Technology Letters}, vol. 14, no. 3, pp.
  355--357, 2002.

\bibitem{Goryachev_054002}
Maxim Goryachev, Warrick~G Farr, Daniel~L Creedon, Yaohui Fan, Mikhail
  Kostylev, and Michael~E Tobar,
\newblock ``High-cooperativity cavity qed with magnons at microwave
  frequencies'',
\newblock {\em Physical Review Applied}, vol. 2, no. 5, pp. 054002, 2014.

\bibitem{Zhang_205003}
Dongshan Zhang, Wenjie Song, and Guozhi Chai,
\newblock ``Spin-wave magnon-polaritons in a split-ring
  resonator/single-crystalline yig system'',
\newblock {\em Journal of Physics D: Applied Physics}, vol. 50, no. 20, pp.
  205003, 2017.

\bibitem{Mathai_054428}
Cijy Mathai, Oleg Shtempluck, and Eyal Buks,
\newblock ``Thermal instability in a ferrimagnetic resonator strongly coupled
  to a loop-gap microwave cavity'',
\newblock {\em Phys. Rev. B}, vol. 104, pp. 054428, Aug 2021.

\bibitem{Nayak_062404}
Banoj~Kumar Nayak, Cijy Mathai, Dekel Meirom, Oleg Shtempluck, and Eyal Buks,
\newblock ``Optical interface for a hybrid magnon--photon resonator'',
\newblock {\em Applied Physics Letters}, vol. 120, no. 6, pp. 062404, 2022.

\bibitem{Wood_1038}
DL~Wood and JP~Remeika,
\newblock ``Effect of impurities on the optical properties of yttrium iron
  garnet'',
\newblock {\em Journal of Applied Physics}, vol. 38, no. 3, pp. 1038--1045,
  1967.

\bibitem{Wang_224410}
Yi-Pu Wang, Guo-Qiang Zhang, Dengke Zhang, Xiao-Qing Luo, Wei Xiong, Shuai-Peng
  Wang, Tie-Fu Li, C-M Hu, and JQ~You,
\newblock ``Magnon kerr effect in a strongly coupled cavity-magnon system'',
\newblock {\em Physical Review B}, vol. 94, no. 22, pp. 224410, 2016.

\bibitem{Mathai_67001}
Cijy Mathai, Sergei Masis, Oleg Shtempluck, Shay Hacohen-Gourgy, and Eyal Buks,
\newblock ``Frequency mixing in a ferrimagnetic sphere resonator'',
\newblock {\em Euro. Phys. Lett.}, vol. 131, 2020.

\bibitem{Freiser_152}
Ml~Freiser,
\newblock ``A survey of magnetooptic effects'',
\newblock {\em IEEE Transactions on magnetics}, vol. 4, no. 2, pp. 152--161,
  1968.

\bibitem{Boardman_197}
Allan~D Boardman and Ming Xie,
\newblock ``Magneto-optics: a critical review'',
\newblock {\em Introduction to Complex Mediums for Optics and
  Electromagnetics}, vol. 123, pp. 197, 2003.

\bibitem{Boardman_388}
Allan~D Boardman and Larry Velasco,
\newblock ``Gyroelectric cubic-quintic dissipative solitons'',
\newblock {\em IEEE Journal of selected topics in quantum electronics}, vol.
  12, no. 3, pp. 388--397, 2006.

\bibitem{Buks_014421}
Eyal Buks and Banoj~Kumar Nayak,
\newblock ``Quantum measurement with recycled photons'',
\newblock {\em Physical Review B}, vol. 105, no. 1, pp. 014421, 2022.

\end{thebibliography}

\end{document}